# A Model of Decision-Making in Groups of Humans

Gabriel Madirolas[1], Alfonso Pérez-Escudero[1], Gonzalo G. de Polavieja[1]
1 Instituto Cajal, Consejo Superior de Investigaciones Científicas, Madrid, Spain

**Abstract**

Decisions by humans depend on their estimations given some uncertain sensory data. These decisions can also be influenced by the behavior of others. Here we present a mathematical model to quantify this influence, inviting a further study on the cognitive consequences of social information. We also expect that the present model can be used for a better understanding of the neural circuits implicated in social processing.

**General model**

We follow a previous paper by Pérez-Escudero & de Polavieja [1], where it was shown that *bayesian estimation* can be assumed in animals' decision-making. We will also take from that paper the concept that each animal effectively decides through *probability matching*, i.e. each option is chosen with a probability given by the process of bayesian estimation.

We are motivated by experiments at which individuals were asked to make a guess about some value or quantity for a certain fact of the real world [2-3]. Let's call $P(X_i | C, B)$ the conditional probability that the correct answer is $X_i$, given social (or public) information $B$, and non-social (or private) information $C$. Here we call social information to whatever poblational or muestral information given to the subjects, such as all the individual previous guesses, or the arithmetic mean of them.

According to *Bayes' Theorem*, the estimated probability that a particular option is the correct is given by

$$P(X_i | C, B) = \frac{P(B | X_i, C) P(X_i | C)}{\sum_{l=1}^{L} P(B | X_l, C) P(X_l | C)}, \qquad (1)$$

where $l$ runs for all the possible guesses, that may be an infinite set. We will focus on the numerator term, because the denominator is just a normalization term, and we are only interested in the shape of the distribution, i.e. the proportion of its value at two particular points.

Let's start with the particular case where the subject is given full information of which guesses have been made by all the previous subjects. If we call $P(X_j | X_i, C)$ the estimated probability that the correct answer is $X_j$, when the actual correct answer is $X_i$, and we assume independence between each other's guesses, we have

$$P(X_i | C, B) \propto P(X_i | C) \prod_{j=1}^{n} P(X_j | X_i, C), \qquad (2)$$

where $n$ is the number of previous guesses. Following the reasoning and data from Lorenz et al. [2], we assume log-normal distributions for all the estimations, as they are justified for variables with high variance with a range of positive values only. We can assume, for the non-social term, a log-normal distribution with median $X_t$:

$$P(X_i | C) = P(X_i | \log-\text{normal}(\log(X_t), \sigma_t)), \qquad (3)$$

and for the estimation of the probability of each previous guess given that $X_i$ is the correct answer, log-normal distributions with median $X_i$ and the same parameter $\sigma$:

$$P(X_j | X_i, C) = P(X_j | \log-\text{normal}(\log(X_i), \sigma)). \qquad (4)$$

Then, expression (2) turns into

$$P(X_i | C, B) \propto P(X_i | \log-\text{normal}(\log(X_t), \sigma_t)) \prod_{j=1}^{n} P(X_j | \log-\text{normal}(\log(X_i), \sigma)). \qquad (5)$$

For the first term of the right side of (5), which is the non-social term of the estimation, we have the following log-normal expression:

$$P(X_i | C) = \frac{1}{x_i \sigma_t \sqrt{2\pi}} e^{-\frac{[\log(x_i) - \log(x_t)]^2}{2\sigma_t^2}}. \qquad (6)$$

For the second term of the right side of (5), which is the social term of the estimation, we have (see proof 1)

$$\prod_{j=1}^{n} \frac{1}{x_j \sigma \sqrt{2\pi}} e^{-\frac{[\log(x_j) - \log(x_i)]^2}{2\sigma^2}} = \frac{1}{\left(\sigma \sqrt{2\pi}\right)^n \prod_{j=1}^{n} x_j} e^{-\frac{n\left[\log(x_i) - \log\left(\left[\prod_{j=1}^{n} x_j\right]^{1/n}\right)\right]^2}{2\sigma^2}}. \qquad (7)$$

Now, as the geometric mean $G$ of a sample is

$$G = \sqrt[n]{\prod_{j=1}^{n} x_j}, \tag{8}$$

and the first multiplicative term of the right side of (7) can be taken as a constant given a set of previous answers, we have for the social term that

$$P(X_i \mid C, B) \propto e^{-\frac{n[\log(x_i) - \log(G)]^2}{2\sigma^2}}. \tag{9}$$

This seems a good result, as the geometric mean is an estimator of the poblational median, which we are assuming is the correct answer for the guessing. So, for the total probability distribution, we have

$$P(X_i \mid C, B) \propto \frac{1}{x_i} e^{-\frac{[\log(x_i) - \log(x_t)]^2}{2\sigma_t^2}} e^{-\frac{n[\log(x_i) - \log(G)]^2}{2\sigma^2}}. \tag{10}$$

If we take

$$\sigma_t^2 = \frac{\sigma^2}{m}, \tag{11}$$

being $m$ a real number to be determined, we have

$$P(X_i \mid C, B) \propto \frac{1}{x_i} e^{-\frac{m[\log(x_i) - \log(x_t)]^2}{2\sigma^2}} e^{-\frac{n[\log(x_i) - \log(G)]^2}{2\sigma^2}}. \tag{12}$$

This can be rewritten as (see proof 2)

$$P(X_i \mid C, B) \propto \frac{1}{x_i} e^{-\frac{(m+n)\left[\log(x_i) - \log([x_t^m G^n]^{\frac{1}{m+n}})\right]^2}{2\sigma^2}} e^{-\frac{mn}{m+n}\frac{[\log(x_t) - \log(G)]^2}{2\sigma^2}}. \tag{13}$$

So far, we have assumed that the non-social term is a log-normal with median $x_t$, for any individual. In this case, the last term in (13) can simply be dropped out, as it is a constant given a correct guess and a particular set of previous guesses. So, for this particular case, we simply have

$$P(X_i \mid C, B) \propto \frac{1}{x_i} e^{-\frac{(m+n)\left[\log(x_i) - \log([x_t^m G^n]^{\frac{1}{m+n}})\right]^2}{2\sigma^2}}. \tag{14}$$

**Proofs**

Proof 1. We are going to use induction to prove:

$$\prod_{j=1}^{n} \frac{1}{x_j \sigma \sqrt{2\pi}} e^{-\frac{[\log(x_j)-\log(x_i)]^2}{2\sigma^2}} = \frac{1}{\left(\sigma \sqrt{2\pi}\right)^n \prod_{j=1}^{n} x_j} e^{-\frac{n\left[\log(x_i)-\log\left(\left[\prod_{j=1}^{n} x_j\right]^{1/n}\right)\right]^2}{2\sigma^2}}, \quad (7)$$

that leads to

$$P(X_i \mid C, B) \propto e^{-\frac{n[\log(x_i)-\log(G)]^2}{2\sigma^2}}. \quad (9)$$

For two terms of the product, we have

$$\frac{1}{x_j \sigma \sqrt{2\pi}} e^{-\frac{[\log(x_j)-\log(x_i)]^2}{2\sigma^2}} \frac{1}{x_k \sigma \sqrt{2\pi}} e^{-\frac{[\log(x_k)-\log(x_i)]^2}{2\sigma^2}} = \frac{1}{x_j x_k \left(\sigma \sqrt{2\pi}\right)^2} e^{-\frac{[\log(x_j)-\log(x_i)]^2+[\log(x_k)-\log(x_i)]^2}{2\sigma^2}}.$$

The numerator of the exponent is

$$\left[\log(x_j)-\log(x_i)\right]^2 + \left[\log(x_k)-\log(x_i)\right]^2 =$$

$$= \log^2(x_j) + \log^2(x_k) + 2\log^2(x_i) - 2\log(x_i)\log(x_j) - 2\log(x_i)\log(x_k) =$$

$$= 2\log^2(x_i) - 2\log x_i(\log x_j + \log x_k) + \log^2(x_j) + \log^2(x_k) =$$

$$= 2\log^2(x_i) - 2\log x_i \log(x_j x_k) + \log^2(x_j x_k) - \log^2(x_j x_k) + \log^2(x_j) + \log^2(x_k) =$$

$$= 2\left(\log^2(x_i) - 2\log x_i \log\left[(x_j x_k)^{1/2}\right] + \log^2\left[(x_j x_k)^{1/2}\right]\right) - \log^2(x_j x_k) + \log^2(x_j) + \log^2(x_k) =$$

$$= 2\left(\log x_i - \log\left[(x_j x_k)^{1/2}\right]\right)^2 - \log^2(x_j x_k) + \log^2(x_j) + \log^2(x_k).$$

If we call

$$A = -\log^2(x_j x_k) + \log^2(x_j) + \log^2(x_k),$$

we have shown that

$$\left[\log(x_j) - \log(x_i)\right]^2 + \left[\log(x_k) - \log(x_i)\right]^2 = 2\left(\log x_i - \log\left[(x_j x_k)^{1/2}\right]\right)^2 + A.$$

Note that, as we are trying to compute the probability distribution for a subject who has been given specific social information, $A$ is a constant for this particular distribution. We then have that

$$\frac{1}{x_j \sigma \sqrt{2\pi}} e^{-\frac{[\log(x_j) - \log(x_i)]^2}{2\sigma^2}} \frac{1}{x_k \sigma \sqrt{2\pi}} e^{-\frac{[\log(x_k) - \log(x_i)]^2}{2\sigma^2}} \propto e^{-\frac{(\log(x_i) - \log[(x_j x_k)^{1/2}])^2}{2\sigma^2}}.$$

Now let's suppose expression (7) proved for $n$ subjects, and consider it written in the form (9). We have to look at

$$e^{-\frac{n[\log(x_i) - \log(G)]^2}{2\sigma^2}} \frac{1}{x_j \sigma \sqrt{2\pi}} e^{-\frac{[\log(x_j) - \log(x_i)]^2}{2\sigma^2}}.$$

To start with, we have

$$n\left[\log(x_i) - \log(G)\right]^2 + \left[\log(x_j) - \log(x_i)\right]^2 =$$

$$= n\log^2(x_i) + n\log^2(G) - 2n\log(x_i)\log(G) + \log^2(x_i) + \log^2(x_j) - 2\log(x_i)\log(x_j) =$$

$$= (n+1)\log^2(x_i) - 2\log(x_i)\log(x_j G^n) + \log^2(x_j) + n\log^2(G) =$$

$$= (n+1)\log^2(x_i) - 2\log(x_i)\log(x_j G^n) + \frac{1}{n+1}\log^2(x_j G^n) - \frac{1}{n+1}\log^2(x_j G^n) + \log^2(x_j) + n\log^2(G) =$$

$$= (n+1)\left(\log^2(x_i) - 2\log(x_i)\log\left[(x_j G^n)^{1/(n+1)}\right] + \log^2\left[(x_j G^n)^{1/(n+1)}\right]\right)$$

$$- \frac{1}{n+1}\log^2(x_j G^n) + \log^2(x_j) + n\log^2(G) =$$

$$= (n+1)\big(\log(x_i) - \log[(x_j G^n)^{1/(n+1)}]\big)^2 - \frac{1}{n+1}\log^2(x_j G^n) + \log^2(x_j) + n\log^2(G).$$

But $(x_j G^n)^{1/(n+1)}$ is the new geometric mean once we have included $x_j$:

$$G' \equiv (x_j G^n)^{1/(n+1)}.$$

Like we have done before, we can define a constant

$$B = -\frac{1}{n+1}\log^2(x_j G^n) + \log^2(x_j) + n\log^2(G),$$

that lets us show that

$$n\big[\log(x_i) - \log(G)\big]^2 + \big[\log(x_j) - \log(x_i)\big]^2 = (n+1)\big(\log(x_i) - \log[(x_j G^n)^{1/(n+1)}]\big)^2 + B.$$

We have proven that

$$e^{-\frac{n[\log(x_i)-\log(G)]^2}{2\sigma^2}} \frac{1}{x_j \sigma\sqrt{2\pi}} e^{-\frac{[\log(x_j)-\log(x_i)]^2}{2\sigma^2}} \propto e^{-\frac{(n+1)(\log(x_i)-\log(G'))}{2\sigma^2}},$$

and so proved (7) and (9).

Proof 2. We are going to prove that (12) is equivalent to (13), i.e.

$$\frac{1}{x_i} e^{-\frac{m[\log(x_i)-\log(x_t)]^2}{2\sigma^2}} e^{-\frac{n[\log(x_i)-\log(G)]^2}{2\sigma^2}} = \frac{1}{x_i} e^{-\frac{(m+n)\left[\log(x_i)-\log([x_t^m G^n]^{\frac{1}{m+n}})\right]^2}{2\sigma^2}} e^{-\frac{\frac{mn}{m+n}[\log(x_t)-\log(G)]^2}{2\sigma^2}}.$$

This will be proved if we show that

$$m\big[\log(x_i) - \log(x_t)\big]^2 + n\big[\log(x_i) - \log(G)\big]^2 =$$

$$= (m+n)\big(\log(x_i) - \log[(x_t^m G^n)^{1/(m+n)}]\big)^2 + \frac{mn}{m+n}\big[\log(x_t) - \log(G)\big]^2.$$

We do it as follows:

$$m\big[\log(x_i) - \log(x_t)\big]^2 + n\big[\log(x_i) - \log(G)\big]^2 =$$

$$= m\log^2(x_i) + m\log^2(x_t) - 2m\log(x_i)\log(x_t) + n\log^2(x_i) + n\log^2(G) - 2n\log(x_i)\log(G) =$$

$$= (m+n)\log^2(x_i) - 2\log(x_i)\log(x_t^m G^n) + m\log^2(x_t) + n\log^2(G) =$$

$$= (m+n)\log^2(x_i) - 2\log(x_i)\log(x_t^m G^n) + \frac{1}{m+n}\log^2(x_t^m G^n) -$$

$$-\frac{1}{m+n}\log^2(x_t^m G^n) + m\log^2(x_t) + n\log^2(G) =$$

$$= (m+n)\left(\log^2(x_i) - 2\log(x_i)\log\left[(x_t^m G^n)^{1/(m+n)}\right] + \log^2\left[(x_t^m G^n)^{1/(m+n)}\right]\right) -$$

$$- (m+n)\log^2\left[(x_t^m G^n)^{1/(m+n)}\right] + m\log^2(x_t) + n\log^2(G) =$$

$$= (m+n)\left(\log^2(x_i) - \log\left[(x_t^m G^n)^{1/(m+n)}\right]\right)^2 + C,$$

where

$$C = -(m+n)\log^2\left[(x_t^m G^n)^{1/(m+n)}\right] + m\log^2(x_t) + n\log^2(G) =$$

$$= -\frac{1}{m+n}\log^2\left[(x_t^m G^n)\right] + m\log^2(x_t) + n\log^2(G) =$$

$$= -\frac{1}{m+n}\left(\log^2(x_t^m) + \log^2(G^n) + 2\log(x_t^m)\log(G^n)\right) + m\log^2(x_t) + n\log^2(G) =$$

$$= \left(-\frac{m^2}{m+n} + m\right)\log^2(x_t) + \left(-\frac{n^2}{m+n} + n\right)\log^2(G) - 2\frac{mn}{m+n}\log(x_t)\log(G) =$$

$$= \frac{mn}{m+n}\left[\log^2(x_t) + \log^2(G) - 2\log(x_t)\log(G)\right] = \frac{mn}{m+n}\left[\log(x_t) - \log(G)\right]^2.$$

This finishes the proof of the equivalence of (12) and (13).